\input iopppt
\pptstyle
\title{Ricci Collineations in Friedmann-Robertson-Walker Spacetimes}

\author{U\~gur Camc\i}
\address{Department of Mathematics, Faculty of Arts and
Sciences, \c{C}anakkale Onsekiz Mart University, 17100
\c{C}anakkale, Turkey \footnote\dag{E-Mail: \ \ {\tt ucamci{\bf
@}comu.edu.tr}}}
\author{Alan Barnes}
\address{Computer Science Group, School of Engineering and Applied Science,
Aston University, Aston Triangle, Birmingham B4 7ET, UK
\footnote\ddag{E-Mail: \ \ {\tt barnesa{\bf @}aston.ac.uk}}}

 \jl{6} \submitted

\date

\beginabstract
Ricci collineations and Ricci inheritance collineations of
Friedmann-Robertson-Walker spacetimes are considered.  When the
Ricci tensor is non-degenerate, it is shown that the spacetime
always admits a fifteen parameter group of Ricci inheritance
collineations; this is the maximal possible dimension for
spacetime manifolds.  The general form of the vector generating
the symmetry is exhibited.  It is also shown, in the generic case,
that the group of Ricci collineations is six-dimensional and
coincides with the isometry group. In special cases the spacetime may
admit either one or four proper Ricci collineations in addition to the
six isometries.  These special 
cases are classified and the general form of the vector fields
generating the Ricci collineations is exhibited. When the Ricci
tensor is degenerate, the groups of Ricci inheritance
collineations and Ricci collineations are infinite-dimensional.
General forms for the generating vectors are obtained.  Similar
results are obtained for matter collineations and matter
inheritance collineations.
\endabstract
\vfill\eject

\section{Introduction}
A vector field $\xi^i$ is said to generate a Ricci inheritance
collineation (Duggal, 1993) if it satisfies the equation ${\cal
L}_\xi R_{ij} = 2\phi R_{ij}$ or equivalently
$$R_{ij,k}\xi^k + R_{ik}\xi^k_{, j}  + R_{jk}\xi^k_{, i} = 2\phi R_{ij}
\eqno(1)$$ If $\phi = 0$, then ${\cal L}_\xi R_{ij} = 0$ and
$\xi^i$ is said to generate a Ricci collineation (Katzin et al.,
1969).   If $\xi^i$ is a Killing vector, then a fortiori  ${\cal
L}_\xi R_{ij} = 0$, thus any isometry is also a Ricci
collineation.  A similar result holds for homothetic Killing
vectors $\xi^i$ for which ${\cal L}_\xi g_{ij} = 2\sigma g_{ij}$
with $\sigma_{,i} = 0$ as hometheties are also affine
collineations and so ${\cal L}_\xi \Gamma^i_{jk} =0$, ${\cal
L}_\xi R^i_{\ jkl} = 0$ and ${\cal L}_\xi R_{ij} = 0$. We will use
the term {\it proper} Ricci collineation to denote a Ricci
collineation which is not an isometry or an homothety.

We note immediately that equation (1), which involves partial
rather than covariant derivatives, has the same form as the
conformal Killing equations but with the metric tensor replaced by
the Ricci tensor. Thus if the Ricci tensor is non-degenerate, that
is has a non-zero determinant, we may apply standard results on
conformal symmetries to deduce that the maximal dimension of the
group of Ricci inheritance collineations (RICs) in a
pseudo-Riemannian manifold of dimension $n$ is $(n+1)(n+2)/2$ and
this occurs if and only if the {\it Ricci tensor metric} is
conformally flat.   By this we mean that the conformal curvature
tensor $^{\rm Ric}C^i_{\ jkl}$ calculated in the standard way but
with $g_{ij}$ replaced everywhere by $R_{ij}$ (including in the
definition of the connection) and with indices raised by the
inverse of $R_{ij}$. Similarly when $\phi = 0$, equation (1) for
Ricci collineations has the same form as Killing's  equations with
$g_{ij}$ again replaced by the $R_{ij}$. If the Ricci tensor is
non-degenerate, we may apply standard results on isometries to
deduce that the maximal dimension of the group of Ricci
collineations (RCs) in a pseudo-Riemannian manifold of dimension
$n$ is $n(n+1)/2$ and this occurs if and only if  the {\it Ricci
tensor metric} has constant curvature.  Thus, for spacetimes, the
maximal dimensions of the groups of RICs and RCs are fifteen and
ten respectively.

The Friedmann-Robertson-Walker (FRW) metric in stereographic
coordinates is
$$ds^2 = dt^2 - S(t)^2(1+K/4 r^2)^{-2}(dx^2 + dy^2 +dz^2) \eqno(2)$$
where $S$ is an arbitrary (non-zero) function of $t$, $K = 0, \pm
1$ and $r^2 = x^2 + y^2 +z^2$.  The Ricci tensor is given by
$$R_{00} =  -3 {\ddot S}/S \qquad R_{\alpha 0} = 0 \qquad
R_{\alpha\beta} = -(2K + 2 {\dot S}^2 +S{\ddot S})/S^2
g_{\alpha\beta} \eqno(3)$$ where Greek indices take the values in
the range $1 \ldots 3$.  The {\it Ricci tensor metric} has the
form
$$ds^2_{\rm Ric} = R_{ij}dx^i dx^j =
A(t)dt^2 + B(t)(1+K/4 r^2)^{-2}(dx^2 + dy^2 +dz^2) \eqno(4)$$
where $A(t) = -3 {\ddot S}/S$ and $B(t) = 2K + 2 {\dot S}^2
+S{\ddot S}$.  Thus the {\it Ricci tensor metric} is of FRW form
with a generalised time coordinate. By rescaling the time
coordinate $d\tilde t = |A|^{1/2} dt$, we could set $A = \pm 1$,
but the new  coordinate would not then be proper time for the
physical metric $g_{ij}$.  In the rest of this section it will be
assumed always that the Ricci tensor is non-degenerate, that is $A
\ne 0$ and $B \ne 0$; the degenerate case will be considered in
section 2. The signature of the Ricci tensor metric depends on the
signs of $A$ and $B$ and is Lorentzian if they have opposite signs
and is positive or negative definite if they have the same sign.
Using Einstein's field equations $A = \kappa/2(\rho + 3 p)$ and $B
=\kappa S^2(\rho-p)/2$ where $\rho$ and $p$ are the fluid density
and pressure respectively. Thus if $\rho > 0$ and $p \ge 0$ as
would normally be assumed, the Ricci tensor is positive definite
if and only if the energy condition $\rho > p$ is satisfied.

For the FRW metric (2), equations (1) for RICs, or equivalently
the conformal Killing equations for the metric (4), become
$$\eqalignno{
B({\tilde g}_{\alpha\beta,\gamma}\xi^\gamma + {\tilde
g}_{\alpha\gamma}\xi^\gamma_{, \beta} + {\tilde
g}_{\beta\gamma}\xi^\gamma_{, \alpha}) &= (2B\phi -\dot
B\xi^0){\tilde g}_{\alpha\beta} &(5a)\cr A\xi^0_{,\alpha} + B
{\tilde g}_{\alpha\beta}\xi^\beta_{,t} &= 0 &(5b)\cr 2 A\xi^0_{,t}
+ \dot A\xi^0 &= 2\phi A &(5c)}$$ where ${\tilde g}_{\alpha\beta}$
is the three-dimensional metric
$$d{\tilde s}^2 = {\tilde g}_{\alpha\beta}dx^\alpha dx^\beta =
(1+K/4 r^2)^{-2}(dx^2 + dy^2 + dz^2) \eqno(6)$$

It is well-known (e.g.~Stephani, 1967) that the FRW metrics are
conformally flat (this conclusion is still true when the signature
is definite rather than Lorentzian). Thus the {\it Ricci tensor
metric} of a FRW spacetime is conformally flat and hence admits a
fifteen parameter group of conformal symmetries or equivalently
the physical metric admits a fifteen parameter group of Ricci
inheritance collineations (RICs).  It is also well-known (Maartens
and Maharaj, 1986) that generic FRW metrics admit only a
six-parameter isometry group, but that in special cases, they
admit either a seven or ten-parameter complete isometry group.
Furthermore for spatially flat FRW metrics the case of a complete
seven-dimensional isometry group is excluded. Again these results
hold when the signature is definite as well as for the Lorentzian
case. When the complete group is seven-dimensional, the spacetime
is a static Einstein universe (or its analogue with definite
signature).  When it is ten-dimensional the spacetime has constant
curvature. Thus we may conclude immediately that group of the
Ricci collineations (RCs) of a FRW spacetime with $K \ne 0$ is
six, seven or ten-dimensional whereas for $K=0$ the complete group
of RCs is either six or ten-dimensional. Dimensions greater than
six occur only when the Ricci tensor metric has constant curvature
or is equivalent to an Einstein static universe.

A priori the possibilty exists that a FRW spacetime always has a
{\it Ricci tensor metric} that admits a seven or ten-dimensional
group of isometries.  If this were the case, then a generic FRW
would admit a seven or ten dimensional group of RCs including one
or four proper RCs respectively.  However, in actual fact for
almost any form for $S(t)$ other than a constant or a constant
multiple of a power, an exponential function, hyperbolic sine and
cosine or trigonometric sine and cosine, the complete group of
Ricci collineations is the six-dimensional isometry group. In
special cases the isometry group of the physical metric can be
six-dimensional whilst the Ricci tensor metric admits a seven or
ten-dimensional isometry group and so proper Ricci collineations
will exist in these cases. These are considered in greater detail
in section 3.

If the isometry group of the physical metric is ten-dimensional,
the spacetime has constant curvature. When the curvature is
non-zero (i.e.~for de Sitter and anti de Sitter spacetimes) the
group of Ricci collineations and isometries coincide and there are
no proper Ricci collineations, whereas for the case of zero
curvature (flat space) any continuous transformation is trivially
a Ricci collineation.  When the physical metric admits a complete
seven-dimensional isometry group ($\dot S = 0$ and $K \ne 0$,
i.e.~the Einstein static universes), the Ricci tensor is
degenerate. Collineations when the Ricci tensor is degenerate are
considered in the next section.

\section{Collineations when the Ricci Tensor is Degenerate}
From equation (3), the Ricci tensor of the FRW metric is degenerate if
$A(t) = -3 {\ddot S}/S = 0$ or $B(t) = 2K + 2 {\dot S}^2 +S{\ddot
S} = 0$ (or both).  If $A=B=0$ then either $S(t)=S_0$ and $K=0$
(Minkowski metric) or $S(t)=t-t_0$ and $K=-1$ (Milne metric),
where $S_0$ and $t_0$ are constants.  The spacetime is flat and so
this case is trivial as the Ricci tensor is zero and any vector
field generates a Ricci collineation.

For the case $B=0$ and $A \ne 0$, equation (5) reveals that the
vector $\xi^i$ generates a RIC if and only if $\xi^0$ is a
function of $t$ only; the spatial components $\xi^\alpha$ are
completely arbitrary. The scale function $\phi = \dot \xi^0 +\dot
A/(2A)\xi^0$ is necessarily a function of $t$ only, but is
otherwise arbitrary.  For RCs $\xi^0$ satisfies the differential
equation $2A \dot \xi^0+ \dot A \xi^0 = 0$. Hence $\xi^0 = d
|A|^{-1/2}$ where $d$ is a constant. The groups of RICs and RCs
are both infinite-dimensional. A first integral of the equation $B
= 2K +2 {\dot S}^2 +S{\ddot S} = 0$ is ${\dot S}^2 = c/S^4 - K$
where $c$ is a constant and hence $\ddot S = -2c/S^5$. For $K=0$
the first integral may be integrated to give $S(t) = S_0
(t-t_0)^{1/3}$ where $S_0$ and $t_0$ are constants. The solution
when $K \ne 0$ involves elliptic integrals (see Appendix A).

If $A = 0$ but $B \ne 0$, then $S = \alpha t +\beta$ and hence $B
= 2(K + \alpha^2)$ where $\alpha$ and $\beta$ are constants where
$\alpha \ne 1$ if $K=-1$ and $\alpha \ne 0$ if $K=0$. Equation (5)
shows that the spatial components $\xi^\alpha$ must be independent
of $t$ and must satisfy the conformal Killing equations of the
three-dimensional metric (6) with conformal factor $\phi$. It is
well known (Robertson and Noonan, 1969) that there are ten
independent conformal Killing vectors of this metric:
$$\eqalignno{
X_1 &= \partial_x &(7a)\cr X_2 &= \partial_y  &(7b)\cr X_3 &=
\partial_z  &(7c)\cr X_4 &= y \partial_z - z \partial_y  &(7d)\cr
X_5 &= z \partial_x - x \partial_z  &(7e)\cr X_6 &= x \partial_y -
y \partial_x  &(7f)\cr X_7 &= x \partial_x + y \partial_y + z
\partial_z  &(7g)\cr X_8 &= (r^2 - 2 x^2)\partial_x - 2 x
y\partial_y -2 x z\partial_z &(7h)\cr X_9 &=  -2 x y\partial_x +
(r^2 - 2 y^2)\partial_y -2 y z\partial_z &(7i)\cr X_{10} &= -2x
z\partial_x -2 y z\partial_y + (r^2 - 2 z^2)\partial_z  &(7j)}
$$
Thus the spatial components $\xi^\alpha$ must have the form
$$\xi^\alpha = \sum_{J=1}^{10} f_J X_J^\alpha \eqno(8)$$
where the $f_J$'s are constants. The associated conformal factor
is given by
$$\sigma = {f_7(1-K/4r^2) -K/2 {\bf f . r} -2 {\bf g .r} \over 1+K/4r^2}
\eqno(9)$$ where ${\bf f} = f_1{\bf i}+ f_2{\bf j} + f_3{\bf k}$,
${\bf g} = f_8{\bf i}+ f_9{\bf j}+f_{10}{\bf k}$ and ${\bf r} =
x{\bf i}+ y{\bf j}+z{\bf k}$; standard 3-dimensional vector
notation has been used for conciseness. The time component $\xi^0$
is completely arbitrary and the scale factor of the RIC is given
by $\phi = \sigma$ where $\sigma$ is given by equation (9). Thus
the group of RICs is infinite dimensional.

For RCs the $\xi^\alpha$'s are independent of $t$ and must satisfy
the Killing equations of the metric (6). As is well-known there
are six independent Killing vectors of this metric, namely:
$$\eqalignno{
Y_1 &= X_1-K/4 X_8 &(10a)\cr Y_2 &= X_2-K/4 X_9 &(10b)\cr Y_3 &=
X_3-K/4 X_{10} &(10c)\cr Y_4 &=X_4 \qquad Y_5 = X_5 \qquad Y_6 =
X_6 &(10d)}$$ where the vectors $X_J$ are as in equation (7). The
most general form for $\xi^\alpha$ is a general linear combination
of these six vectors. The time component $\xi^0$ is completely
arbitrary. and thus the group of RCs is infinite-dimensional.

If $\alpha = 0$ the spacetime is a static Einstein universe. If
$\alpha \ne 0$, by a translation of the time coordinate we may set
$\beta = 0$ and hence $S = \alpha t$.  We note that the physical
metric admits a homothetic Killing vector given by $Z =
t\partial_t$ with associated conformal factor $\sigma = 1$
(Maartens and Maharaj, 1986). These results on Ricci collineations
generalise those of Green et al.~(1977) and N\'u\~nez et
al.~(1990) who investigated Ricci collineations in which the
spatial components $\xi^\alpha$ of the generating vector vanished
or were purely radial respectively.

\section{The Generating Vectors of RICs}
In this section an expression is presented for the general form of
a vector generating a Ricci inheritance collineation in a FRW
spacetime when the Ricci tensor is non-degenerate.  Since Ricci
inheritance collineations are conformal symmetries of the Ricci
tensor metric which is of FRW form, essentially the results are
already in the literature.  The isometries and conformal
symmetries of FRW metrics have been extensively studied by
Maartens and Maharaj (1986) and by Keane and Barrett (2000) and
only need to be reinterpreted in the current context.  However, in
both these papers the authors used curvature coordinates (rather
than stereographic coordinates) for the spatial metric and
naturally proper time was used as a coordinate. Furthermore only
metrics with Lorentzian signature were considered.  In the current
context since proper time of the physical metric will not usually
coincide with `proper time' of the Ricci tensor metric, it is
convenient to work with the metric in the form of equation (4)
without specialising the time coordinate and allowing both
Lorentzian and definite signatures.  Stereographic coordinates are
used rather than curvature coordinates as this simplifies the form
of the results slightly. Essentially the approach used is to
`lift' the ten conformal Killing vectors of the hypersurfaces of
constant $t$ to become conformal Killing vectors of the whole
spacetime.

The spatial components $\xi^\alpha$ of any conformal Killing
vector of the Ricci tensor metric (4) satisfy the conformal
Killing equations for the metric (6) and thus must have the form
$$\xi^\alpha = \sum_{J=1}^{10} f_J(t) X_J^\alpha \eqno(11)$$
where $X_J$ for $J=1 \ldots 10$ are given by equation (6) and the
$f_J$'s are functions of $t$.  The corresponding conformal factor
is given by
$$\phi = {f_7(1-K/4r^2) -K/2 {\bf f . r} -2 {\bf g .r} \over 1+K/4 r^2}
+ {\dot B \xi^0 \over 2B} \eqno(12)$$ where the three-dimensional
vector notation of section 2 has again be used. Equation (5b)
leads to the following compatibility conditions
$${\dot f}_8 = K/4 {\dot f}_1 \qquad  {\dot f}_9 = K/4 {\dot f}_2
\qquad {\dot f}_{10} = K/4 {\dot f}_3 \qquad {\dot f}_4 = {\dot
f}_5  = {\dot f}_6 =0 \eqno(13)$$

For $K=0$, integration of equations (5b) and (5c) gives
$$\xi^0 = -B/A({\bf {\dot f} . r} +1/2 r^2{\dot f}_7) + h(t) \eqno(14)$$
where the ${\bf f}(t)$ and $h(t)$ satisfy the following
compatibility conditions:
$$\eqalignno{
{\ddot {\bf f}} &=1/2(\dot A/A -\dot B/B){\dot {\bf f}} + 2A/B{\bf
g} &(15)\cr {\ddot f}_7 &=1/2(\dot A/A -\dot B/B){\dot f}_7
&(16)\cr {\dot h} &= -1/2(\dot A/A -\dot B/B)h + f_7 &(17)}
$$

Equations (13) and (15)-(17) constitute a linear differential
system for eleven unknowns $f_J$ and $h$. Introducing new unknowns
for the first derivatives of $f_1$, $f_2$, $f_3$ and $f_7$ we may
reduce it to a first order linear differential system for fifteen
unknowns. The general solution thus depends on 15 arbitrary
constants and hence (as expected) there are 15 independent
conformal Killing vectors of the Ricci tensor metric or
equivalently 15 RICs of the physical metric. The general solution
of the system is given by
$$\eqalignno{
f_i(t) &= a_i T + b_i + \epsilon c_i T^2 &(18a)\cr f_{3+i}(t)& =
d_i &(18b)\cr f_7(t) &= kT + l &(18c)\cr f_{7+i}(t) &= c_i
&(18d)\cr h(t) &= |B/A|^{1/2}(1/2 k T^2 + l T + m) &(18e)}$$ where
$k$, $l$, $m$, $a_i$, $b_i$, $c_i$ and $d_i$ ($i =1 \ldots 3$) are
arbitrary constants; the constant $\epsilon$ is $+1$ or $-1$ when
the Ricci tensor metric has definite or Lorentzian signature
respectively.  The quantity $T$ is given by
$$T = \int |A/B|^{1/2} dt \eqno(19)$$
and is a `conformal time' coordinate for the Ricci tensor metric.
For $K=0$, the spatial components $\xi^\alpha$ of any vector
generating an RIC are given by equation (11) with the $f_J(t)$'s
given by equation (18) and the time component $\xi^0$ is given by
$$\xi^0 = |B/A|^{1/2}( 1/2k T^2 + lT + m
-\epsilon {\bf a.r} - 2T {\bf c.r} - 1/2\epsilon kr^2) \eqno(20)$$

For $K \ne 0$ the corresponding results are
$$\xi^0 = -B/A({\bf {\dot f} . r} -2/K {\dot f}_7)/(1+K/4 r^2) + h(t)
\eqno(21)$$ and
$$\eqalignno{
{\ddot {\bf f}} &=1/2(\dot A/A -\dot B/B){\dot {\bf f}} + \epsilon
A/B(K/2 {\bf f} + 2{\bf g})  &(22)\cr {\ddot f}_7 &=1/2(\dot A/A
-\dot B/B){\dot f}_7 + \epsilon K A/B f_7 &(23)\cr {\dot h} &=
-1/2(\dot A/A -\dot B/B)h - f_7 &(24)}
$$
The linear system of differential equations (13) and (22)-(24)
again has a fifteen parameter solution.  When $K\epsilon = -1$,
$$\eqalignno{
f_i(t) &= a_i \cos(T) +b_i \sin(T) -2Kc_i &(25a)\cr f_{3+i}(t) &=
d_i &(25b)\cr f_7(t) &= k \cos(T) + l\sin(T) &(25c)\cr f_{7+i} &=
K/4(a_i \cos(T) +b_i \sin(T)) +1/2c_i &(25d)\cr h(t) &=
|B/A|^{1/2}( -k \sin(T) +l \cos(T) + m) &(25e)}$$ where $T$,
$\epsilon$ etc.~are as in the $K=0$ case. The spatial components
$\xi^\alpha$ of any vector generating an RIC are given by equation
(11) with the $f_J(t)$'s given by equation (25) and the time
component $\xi^0$ is given by
$$\fl \xi^0 =
|B/A|^{1/2}\Bigl({\epsilon{\bf a.r}\sin(T) -\epsilon{\bf
b.r}\cos(T) +2k\sin(T) -2l\cos(T) \over 1+K/4r^2} \cr \lo
-k\sin(T) + l\cos(T) +m\Bigr) \qquad\qquad {\rm for \ } K\epsilon
=-1 \eqno(26)$$ When $K\epsilon = +1$,
$$\eqalignno{
f_i(t) &= a_i \cosh(T) +b_i \sinh(T) -2Kc_i &(27a)\cr f_{3+i}(t)
&= d_i &(27b)\cr f_7(t) &= k \cosh(T) + l\sinh(T) &(27c)\cr
f_{7+i} &= K/4(a_i \cosh(T) +b_i \sinh(T)) +1/2c_i &(27d)\cr h(t)
&= |B/A|^{1/2}( -k \sinh(T) - l \cosh(T) + m) &(27e)}$$ The
spatial components $\xi^\alpha$ are given by equation (11) with
the $f_J(t)$'s given by equation (27) and the time component
$\xi^0$ is
$$\fl \xi^0 =
|B/A|^{1/2}\Bigl({-\epsilon{\bf a.r}\sinh(T) -\epsilon{\bf
b.r}\cosh(T) +2k\sinh(T) + 2l\cosh(T) \over 1+K/4r^2}\cr \lo -
k\sinh(T) - l\cosh(T) +m\Bigr) \qquad \qquad {\rm for\ }
K\epsilon=+1 \eqno(28)$$

\section{Proper Ricci Collineations in the Non-degenerate Case}
In this section we assume that the Ricci tensor of the FRW metric
is not degenerate. We investigate the conditions on the scale
factor $S(t)$ such that the spacetime admits proper RCs in
addition to the six Killing vectors in equation (10). Thus by the
discussion of section 1 we need to search for extra isometries of
the Ricci tensor metric (4). The only possibilities are one extra
Killing vector or four extra Killing vectors (Maartens and
Maharaj, 1986).

When there is only one extra Killing vector, $K \ne 0$ and the
Ricci tensor metric must be equivalent to the metric of an
Einstein static universe (or its analogue with definite signature)
and thus $\dot B = 0$. The FRW scale factor $S$ and the vector $Z$
generating the proper RC satisfy
$$\dot S^2 = d + c/S^4 \qquad \qquad Z = S^3\partial_t \eqno(29)$$
where $c$ and $d$ are constants. The functions in the Ricci tensor
metric satisfy $B = 2(K+d)$ and $A = 6c/S^6$ and hence for
non-degeneracy: $d \ne -K$ and $c \ne 0$. When $d=0$, the scale
factor has the form $S(t) = S_0 (t-t_0)^{1/3}$ where $S_0$ and
$t_0$ are constants, but in general $S(t)$ will involve elliptic
integrals (see Appendix A).

When there are four extra isometries the Ricci tensor metric has
constant curvature; this condition is equivalent to
$$\ddot B -\dot B(\dot B/B + \dot A/(2A)) + 2K A = 0 \eqno(30)$$
Eliminating $A$ and $B$ using equation (3), a complicated fourth
order differential equation for $S(t)$ results\footnote\dag{see
equation {\tt rc10} in the Reduce output file with URL: {\tt
http://www.aston.ac.uk/{\~ \null}barnesa/cc.res}}. To date we have not
succeeded in analysing this fully. However when $K=0$, one
solution is $S(t) = c (t-t_0)^d$ where $c$, $d$ and $t_0$ are
constants. When $d \ne 1$, this physical metric has only six
Killing vectors plus a homothetic vector
$$Z = (t-t_0)/(1-d)\partial_t + x\partial_x + y\partial_y + z\partial_z
\eqno(31)$$ with corresponding conformal factor $\sigma =
1/(1-d)$. However it has three proper RCs plus seven RCs generated
by the homothetic vector $Z$ and the six spatial Killing vectors.
This example and the ones given in section 2 are the only FRW
metrics admitting an homothetic vector (Maartens and Maharaj,
1986).

Other solutions of equation (30) are known:
$$\eqalignno{
S(t) &= c e^{t/d}  &{\rm for\ } K = 0 \qquad \qquad (32a)\cr S(t)
&=  d \cosh((t-t_0)/d) &{\rm for\ } K = 1 \qquad \qquad (32b)\cr
S(t) &=  d \sinh((t-t_0)/d) &{\rm for\ } K = -1 \qquad \qquad
(32c)\cr S(t) &=  d \sin((t-t_0)/d) &{\rm for\ } K = -1 \qquad
\qquad (32d)}$$ However these all describe spacetimes of constant
curvature and so the group of RCs coincides with the isometry
group.

As we have seen above generic FRW metrics do not admit proper
Ricci collineations; these exist only in special cases when the
scale factor $S(t)$ has a special form such that the {Ricci tensor
metric} is degenerate, or of constant curvature or isometric to
the metric of a static Einstein  universe. This result contradicts
a claim by Carot et al.~(1997) that all FRW metrics admit a RC
with a generating vector $\xi^i$ which has only $t$ and $r$
components; we find that the vector given by their equation (75)
only satisfies two of the four required relations given in their
equation (74).  According to our calculations the radial component
can only be non-zero if $K=0$ and if the scale factor $S(t)$
satisfies equation (30). For $K \ne 0$, the radial component must
vanish and the scale factor must satisfy equation (29). Similar
results have been derived under much more restrictive assumptions
by N\'u\~nez et al.~(1990).

The analysis above does not directly yield expressions for the
additional Killing vectors of the Ricci tensor metric when these
exist. However, in section 3 expressions for the most general
vector $\xi^i$ generating a RIC were obtained. For RCs we must
impose the condition $\phi = 0$ where $\phi$ is given by equation
(12). Using equations (18) and (19) we see that $\phi=0$ is a
second order polynomial equation in $x$, $y$ and $z$ with
coefficients which are functions of $t$ only. Equating these
coefficients to zero leads to 4 conditions restricting $B$ and the
constants $a_i$, $c_i$, $k$, $l$ and $m$.

For $K=0$, equating the coefficient of $r^2$ to zero we find
$k\dot B = 0$.  Now if $\dot B =0$, it follows that
$$f_i = a_iT+b_i \qquad f_{i+3} = d_i \qquad f_7=f_{7+i}=0
\qquad h =|B/A|^{1/2}m \eqno(33)$$ Thus the general vector
generating an RC involves ten arbitrary constants; it has spatial
components  $\xi^\alpha$ given by equation (11) with the $f_j$'s
given by equation (33) and time component given by  $\xi^0 =
|B/A|^{1/2}(m -\epsilon {\bf a.r})$.  There are 4 proper RCs plus
the usual six isometries. If $\dot B \ne 0$, then $k=0$. The
remaining equations are compatible only if $B =B_0/(T-T_0)^2$,
where $B_0$ and $T_0$ are constants. However $T$ is only defined
up to an additive constant and so, without loss of generality, we
may set $T_0 = 0$. It follows that
$$f_i = b_i \qquad f_{i+3} = d_i \qquad f_7=l \qquad f_{7+i}=c_i
\qquad h =|B/A|^{1/2} lT \eqno(34)$$ Thus the general vector
generating an RC involves ten arbitrary constants; its spatial
components $\xi^\alpha$ are given by equation (8) with the $f_j$'s
given by equation (34) and its time component is $\xi^0 =
|B/A|^{1/2}T(l - 2{\bf c.r})$.  Again there are 4 proper RCs plus
the usual six isometries. The metric with scale factor $S(t) =
ct^d$  discussed earlier in this section belongs to this class.
Similar results (in curvature coordinates) have recently been
obtained independently by Apostolopoulos and Tsamparlis (2001) for
the spatially flat case ($K=0$) only.

For $K \ne 0$, the condition $\phi = 0$ leads to the following
relations
$$\eqalignno{
&\dot B h -2B f_7 = 0 &(35a)\cr &2KA f_7 +\dot B \dot f_7 = 0
&(35b)\cr &A(K{\bf f} +4{\bf g}) + \dot B {\bf \dot f} = 0
&(35c)}$$ If $\dot B = 0$, it follows that $a_i=b_i=k=l=0$ and
$h(t) = m |B/A|^{1/2}$.  Thus there is only one proper RC
generated by the vector $Z = |B/A|^{1/2} \partial_t$. If $\dot B
\ne 0$, the equations are only compatible for special forms of
$B$. When $K\epsilon = -1$, essentially the only possibility is
$B=B_0/ \sin^2(T)$. Then $b_i = l = m =0$ and the general vector
generating an RC involves ten arbitrary constants $a_i$, $c_i$,
$d_i$ and $k$ and is given by equations (8) and (26) with the
$f_J$'s given by equation (25).  Again there are four proper RCs
plus the usual six isometries. The case $K\epsilon = +1$ is
similar except that now there are essentially three possibilities:
$$B=B_0/\cosh^2(T)\qquad B=B_0/\sinh^2(T)\qquad B=\exp(\pm 2T)
\eqno(36)$$ In each case there is a ten-parameter group of RCs
consisting of four proper RCs and the usual six isometries. For
example when the first of equations (36) holds, $a_i = k = m= 0$
and the general vector generating an RC involves ten arbitrary
constants $b_i$, $c_i$, $d_i$ and $l$ and is given by equation (8)
and (28) with the $f_J$'s given by equation (27).  The other two
cases are similar.

The analysis of sections 3 and 4 involves some rather heavy
algebraic manipulations and has been checked using the computer
algebra systems CLASSI ({\AA}man, 1987) and Reduce (Hearn, 1995).

\section{Matter Collineations}
In this section we consider briefly matter inheritance
collineations and matter collineations admitted by FRW spacetimes.
As we will assume the validity of Einstein's field equations
$G_{ij} = \kappa T_{ij}$, a vector $\xi^i$ generates a matter
inheritance collineation if ${\cal L}_\xi G_{ij} = 2\phi G_{ij}$.  If
$\phi = 0$, it is said to admit a matter collineation. If the Einstein
tensor is non-degenerate, then following the same arguments as in
the Introduction, we may conclude that the maximal groups of
matter inheritance collineations and matter collineations are
fifteen and ten respectively. Matter inheritance collineations and
plain matter collineations are respectively conformal symmetries
and isometries of the {\it Einstein tensor metric}.

For FRW spacetimes the {\it Einstein tensor metric} has FRW form:
$$ds_{Ein}^2 = G_{ij}dx^i dx^j = A(t) dt^2
+ B(t)(1+K/4 r^2)^{-2}(dx^2 + dy^2 +dz^2) \eqno(37)$$ where now
$$A=3(K+\dot S^2)/S^2 = \kappa \rho \qquad
B = -(K +\dot S^2 + 2 S \ddot S)= \kappa p S^2 \eqno(38)$$ In the
rest of this section $A$ and $B$ will refer to those quantities
defined in equation (38) rather than those defined immediately
after equation (4).

The Einstein tensor metric is positive-definite when $\rho > 0$
and $p > 0$.  Only two distinct degenerate cases occur: $A=B=0$
and $B=0$ with $A \ne 0$ as the condition $A=0$ implies $B=0$. In
the first case the Einstein tensor vanishes and the spacetime is
flat.  Trivially any continuous transformation is a matter
(inheritance) collineation. The second case corresponds to zero
pressure (or matter-dominated) FRW models and in this case the
groups of matter inheritance collineations and matter
collineations are both infinite-dimensional. The generators have
spatial components $\xi^\alpha$ which are completely arbitrary and
the time component $\xi^0$ is a function of $t$ only; for
inheritance collineations $\xi^0$ is otherwise arbitrary, whereas
for plain matter collineations it is given by $\xi^0 =
d|A|^{-1/2}$. The form of the scale factor $S(t)$ for
pressure-free FRW models is well known and appears in many books
on relativity and cosmology, for example Stephani (1982).

In the non-degenerate case the situation is again similar to, but
slightly simpler than, the case of Ricci collineations discussed
in sections 1, 3 and 4. There is always a fifteen parameter group
of matter inheritance collineations which are the conformal
symmetries of the Einstein tensor metric (which is confromally
flat). The generating vectors take the same form as in section 3.
For generic FRW metrics the physical metric and the Einstein
tensor metric will admit the same isometry group; the group of
matter collineations coincides with the usual six-dimensional
group of motions generated by the Killing vectors given in
equation (10).  If the physical metric admits a (complete)
isometry group of dimension seven, that is if the spacetime is an
Einstein static universe ($k \ne 0$ and $\dot S = 0$), then the
Einstein tensor metric is also isometric to that of an Einstein
static universe.  The functions $A$ and $B$ are both constant and
the group of matter collineations coincides with the isometry
group.  If the physical metric has (non-zero) constant curvature,
the group of matter collineations again coincides with the
isometry group which is now ten-dimensional.

In all the non-degenerate cases discussed above there are no
proper matter collineations, but the possibility remains that the
physical metric admits only a six-dimensional isometry group
whereas the Einstein tensor metric admits a higher dimensional
group of motions (necessarily of dimension 7 or 10) and so admits
matter collineations which are not isometries. As in section 4 two
cases arise.

If $K \ne 0$, $\dot S \ne 0$ but $\dot B = 0$, the Einstein tensor
metric is static whereas the physical metric is not. The group of
matter collineations is seven-dimensional and is generated by the
usual six Killing vectors plus the vector $Z=
|A|^{-1/2}\partial_t$ The scale factor $S(t)$ satisfies $2S\ddot S
+\dot S^2 = -(K+B_0)$ where $B_0$ is a constant ($=B$).  This is
the same relation as that satisfied by the scale factor for a
pressure-free FRW model, but with $K$ replaced by $K+B_0$. Thus
the integrated forms for $S(t)$ can easily be obtained from the
standard forms in the literature (Stephani, 1982).

If the Einstein tensor metric has constant curvature, but the
physical metric does not, then the group of matter collineations
will be ten-dimensional and hence there will be four independent
proper matter collineations in addition to the usual six Killing
vectors.  The functions $A$ and $B$ appearing in equations (37)
and (38) must satisfy equation (30) which again leads to a
complicated fourth order non-linear differential expression for
the scale factor $S(t)$\footnote\ddag{see equation {\tt mc10} in
the Reduce output file with URL: {\tt
http://www.aston.ac.uk/{\~ \null}barnesa/cc.res}}. Again we have not been
able to compeletly analyse this expression; however when $K=0$,
$S(t) = c(t-t_0)^d$ is again a solution. Thus the metric admits
three proper matter collineations plus seven matter collineations
generated by the homothetic vector $Z$ in equation (31) and the
six Killing vectors. The scale factors $S(t)$ given by equation
(32), all of which correspond to physical metrics of constant
curvature again satisfy this fourth order differential expression.

The form of the generating vectors can be obtained using the
methods of section 4. We conclude by briefly investigating the
relation between Ricci and matter collineations. Suppose $\xi^i$
satisfies ${\cal L}_\xi R_{ij} = 2\phi R_{ij}$ and hence generates
a Ricci inheritance collineation.  If $R=0$, a Ricci (inheritance)
collineation is obviously a matter (inheritance) collineation and
vice-versa. A simple calculation shows that
$${\cal L}_\xi G_{ij} = 2\phi G_{ij} +1/2(h^{kl}R_{kl}g_{ij} -R h_{ij})
\eqno(39)$$ where $h_{ij} = {\cal L}_\xi g_{ij}$. Hence if $R \ne
0$, a Ricci collineation is a matter collineation iff $h_{ij} =
{\cal L}_\xi g_{ij} \propto g_{ij}$; that is iff $\xi^i$ is a
conformal Killing vector.  A similar results holds for inheritance
collineations (with the same `inheriting' factor $\phi$).

\section{Summary}
When the Ricci tensor is degenerate, the groups of Ricci
collineations and Ricci inheritance collineations of a FRW
spacetime are both infinite-dimensional. The FRW scale factors
leading to  these infinite-dimensional groups have been obtained
in closed form. When the Ricci tensor is non-degenerate, Ricci
collineations and Ricci inheritance collineations are respectively
isometries and conformal symmetries of the {\it Ricci tensor
metric}. For FRW metrics the Ricci tensor metric also assumes FRW
form and so it follows immediately that the group of Ricci
inheritance collineations is of dimension fifteen and, for a
generic FRW metric, the group of Ricci collineations is of
dimension six and coincides with the isometry group. For special
FRW metrics the group of Ricci collineations may be larger; it has
dimension seven when the Ricci tensor metric is isometric to that
of an Einstein static universe and  dimension ten when the Ricci
tensor metric has constant curvature. Conditions on the scale
factor of the physical metric for these higher dimensional groups
to occur have been obtained. Those scale factors leading to a
seven-dimensional group are known explicitly whereas those leading
to a ten-dimensional group are currently only known up to the
solution of a complicated non-linear fourth order differential
equation.  A few special solutions of this equation have, however,
been obtained.

The most general form of a vector generating a Ricci collineation
or Ricci inheritance collineation has been obtained for both the
degenerate and non-degenerate cases. A parallel set of results has
been obtained for matter collineations and matter inheritance
collineations of FRW metrics.

\ackn One of us (UC) would like to thank T\"{U}B\.{I}TAK-Feza
G\"{u}rsey Institute, \.{I}stanbul, for the hospitality he
received during his stay in the summer term of 2001.

\appendix{A}{Solution of the Equation $\dot{B}(t) = 0$}
If the function $B(t)$  of section 4 is a constant, $B_0$ say,
then the scale factor $S(t)$ satisfies $S\ddot S + 2 \dot S^2 = -2
(K - B_0/2)$. Integration of this equation yields
$$ t - t_0 = 3 \int{\frac{S^2 dS}{\left( S_0 - 9(K-B_0/2)
S^4\right)^{1/2}}} \eqno(A1) $$ where $t_0$ and $S_0$ are
integration constants. It is possible to express the integral (A1)
for all values of $K$ in terms of hypergeometric functions; it is
given by the following unified form
$$ t - t_0 = \frac{S^3}{S_0^{1/2}} \,\, _2 F _1 \left( \frac{1}{2},
\frac{3}{4};\frac{7}{4};\frac{18K-B_0}{2S_0}S^4 \right),
\eqno(A2)$$ where $_2 F_1$ is a hypergeometric function and $S_0
\neq 0$. The above unified form clearly depends on the values of
$S_0$ and $K$.

In the special case $B(t) = 0$ given in section 2, the constant
$B_0$ vanishes.  Then $\rho = p$ and we have a fluid with the
equation of state of `stiff' matter. In this case, integrating the
equation $B(t) = 0$, three kinds of solution for $S(t)$ occur: (i)
$K = 0$, (ii) $K = +1$ and (iii) $K = -1$. In case (i) the
integral is elementary, namely $S = S_0^{1/6} (t-t_0)^{1/3}$. For
the other two cases, in addition to the solution obtained by
N\'u\~nez et al.~(1990), the closed form of the solution of the
integral (A1) may be found in terms of elliptic integrals of the
second kind as follows:
$$ t -t_0 = \left( \frac{S_0}{729 K^3} \right)^{1/4} \left[ E
\left( \frac{\pi}{4},2\right) - E \left(\gamma , 2\right) \right],
\eqno(A3)$$ where $\gamma = (1/2)\,cos^{-1} \left( \frac{3
S^2}{\sqrt{S_0/K}} \right)$.

\references \refbk{Apostolopoulos P S and Tsamparlis M
2001}{Preprint: Comment on Ricci Collineations for spherically
symmetric space-times}{arXiv:gr-qc/0108064} \refjl{Carot J,
N\'u\~nez L A and Percoco U 1997}{\GRG}{29}{1223--37}
\refjl{Duggal K L 1993}{Acta Appl. Math.}{31}{225} \refjl{Green L
H, Norris L K, Oliver D R and Davis W R  1977}{\GRG} {8}{731}
\refbk{Hearn A C 1995}{Reduce User's Manual, Version 3.6}{Rand,
Santa Monica, CA} \refjl{Katzin G H, Levine J and Davis H R
1969}{\JMP}{10}{617--9} \refjl{Keane A J and Barrett R K
2000}{\CQG}{17}{201--18} \refjl{Maartens R and Maharaj S D
1986}{\CQG}{3}{1005--11} \refjl{N\'u\~nez L A, Percoco U and
Villalba V M  1990}{\JMP}{31}{137--9} \refbk{Robertson H P and
Noonan T W  1969}{Relativity and Cosmology}{Sanders, Philadelphia
PA} \refjl{Stephani H 1967}{Commun. Math. Phys.}{4}{137--42}
\refbk{Stephani H 1982}{General Relativity}{Cambridge University
Press, Cambridge UK} \refbk{{\AA}man J E 1987}{Manual for CLASSI:
classification program for geometries in general
relativity}{University of Stockholm, Institute of Theoretical
Physics  technical report} \bye